\documentstyle[12pt]{article}
\input epsf
\font\upright=cmu10 scaled\magstep1

\newcommand{\R}{\hbox{\upright\rlap{I}\kern 1.7pt R}}
\newcommand{\C}{\hbox{\upright\rlap{I}\kern -1.9pt C}}
\newcommand{\la}{\label}

\newcommand{\be}{\begin{equation}}
\newcommand{\ee}{\end{equation}}
\newcommand{\ba}{\begin{eqnarray}}
\newcommand{\ea}{\end{eqnarray}}
\newcommand{\bastar}{\begin{eqnarray*}}
\newcommand{\eastar}{\end{eqnarray*}}
\begin{document}

\begin{titlepage}

\begin{center}
\vskip 4.0cm
{\bf \large ALGEBRAIC SHAPE INVARIANT POTENTIALS AS \\
THE GENERALIZED DEFORMED OSCILLATOR \\}
\end{center}

\vskip 2.0cm

\begin{center}
{\bf Wang-Chang Su$^{*}$ }  \\
\vskip 0.3cm
{\it Department of Physics \\
     National Chung-Cheng University \\
     Chia-Yi 621, Taiwan } \\
\end{center}

\vskip 0.5cm

{\rm
Within the framework of supersymmetric quantum mechanics, we study the simplified version of potential algebra of shape invariance condition in $k$ steps, where $k$ is an arbitrary positive integer.
The associated potential algebra is found to be equivalent to the generalized deformed oscillator algebra that has a built-in $Z_{k}$-grading structure.
The algebraic realization of shape invariance condition in $k$ steps is therefore formulated by the method of $Z_k$-graded deformed oscillator.
Based on this formulation, we explicitly construct the general algebraic properties for shape invariant potentials in $k$ steps, in which the parameters of partner potentials are related to each other by translation \( a_1=a_0+\delta \).
The obtained results include the cyclic shape invariant potentials of period $k$ as a special case.
}

\noindent\vfill

\begin{flushleft}
\rule{5.1 in}{.007 in} \\
$^{*}$ \hskip 0.2cm {\small  E-mail: \bf suw@phy.ccu.edu.tw } \\
\end{flushleft}

\end{titlepage}


\section{\la{Introduction}
             Introduction}

Supersymmetry (SUSY) is the symmetry between bosonic and fermionic degrees of freedom.
The idea of SUSY was initially introduced to solve hierarchy problem in grand unified theories.
SUSY is not an exact symmetry of Nature, therefore it has to be broken at some stages.
It is however difficult to determine whether SUSY is broken or not in supersymmetric quantum field theories.
Supersymmetric quantum mechanics (SUSYQM) as a result becomes a testing ground to understand non-perturbatively SUSY breaking \cite{witten,cf}.
For a review of SUSYQM, please refer to Ref. \cite{cks} and references therein.

In SUSYQM, one construct the SUSY partner Hamiltonian, starting from a given one-dimensional Hamiltonian by the method of factorization \cite{ih}.
The process can be used successively to generate an entire hierarchy of the isospectral Hamiltonians.
Let us be more specific.
Two potentials $V^{(-)}(x,a_0)$ and $V^{(+)}(x,a_0)$ are said to be SUSY partner potentials, if they are related to the superpotential \( W(x,a_0) \) by
\be
V^{(\pm)} (x,a_0) \, = \, W^2(x,a_0) \pm W'(x,a_0) \, ,
\la{Vpm}
\ee
where \( W'(x,a_0) \equiv \frac{{\rm d}}{{\rm d}x} W(x,a_0) \) and $a_0$ is a set of parameters.
In units of \( \hbar = 2m = 1 \), the associated SUSY partner Hamiltonians $H^{(-)} (x,a_0)$ and $H^{(+)} (x,a_0)$ have the standard forms
\be
H^{(\pm)} (x,a_0) \, = \, -\frac{{\rm d}^2}{{\rm d}x^2} + V^{(\pm)}(x,a_0) \, .
\la{Hpm}
\ee
In the case of unbroken SUSY, a special property regarding SUSYQM is that except for the zero-energy eigenstate, the SUSY partner Hamiltonians $H^{(\pm)} (x,a_0)$ are found to be exactly isospectral.
Further, if we know all the eigenstates of $H^{(-)}(x,a_0)$ we can determine all the eigenstates of $H^{(+)}(x,a_0)$, and vice versa, except for the zero-energy ground-state eigenfunction.

SUSYQM has been shown to provide a key ingredient to explore exactly solvable potentials for Schr\"odinger equation in nonrelativistic quantum mechanics.
In this respect, the concept of shape invariance condition \cite{gendenshtein} in the formalism of SUSYQM becomes very critical, because it leads immediately to an integrability condition to the problem.
What we mean by shape invariance is that the pair of partner potentials $V^{(\pm)} (x,a_0)$ defined in Eq. (\ref{Vpm}) are similar in shape but differ only up to a change of parameters and additive constants.
Mathematically, the condition reads\footnote{Since the partner potentials $V^{(-)}(x,a_1)$ and $V^{(+)}(x,a_0)$ are related to each other by one relation, Eq. (\ref{SIP1}) is thus called one-step shape invariance condition.}
\be
V^{(+)} (x,a_0) \, = \, V^{(-)} (x,a_1) + R(a_0) \, ,
\la{SIP1}
\ee
where $a_1 = f(a_0)$ is a function of $a_0$ and the remainder $R(a_0)$ is independent of $x$.
By Eq. (\ref{SIP1}), the entire spectrum of eigenenergies for the initial Hamiltonian $H^{(-)}(x,a_0)$ (\ref{Hpm}) can be obtained algebraically as ($n=1,2,3,\cdots$) \cite{gendenshtein,ddck}
\be
E^{(-)}_0 = 0 \, , ~~~~~ E^{(-)}_n = \sum_{i=0}^{n-1} R(a_i) \, .
\ee
Here, we assume that the superpotential $W(x,a_0)$ is constructed in such a way that the Hamiltonian $H^{(-)}(x,a_0)$ possesses the unique zero-energy ground state.
Many interesting classes of solvable shape invariant potentials in one-step that retain SUSY have been reported and discussed \cite{cgk}-\cite{bdgkps}, including all the analytically solvable potentials known in the context of nonrelativistic quantum mechanics.

To obtain more solvable shape invariant potentials, the concept of shape invariance can be extended to two and even multi-steps \cite{bdgkps}.
Based on this method, various shape invariant potentials in two or higher steps have been obtained
\cite{singular}-\cite{su1}.
The extension to shape invariance in multi-steps is rather straightforward.
So, let us consider the case of shape invariance condition in $k$ steps with unbroken SUSY, where $k$ is an arbitrary positive integer.
We are given $k$ superpotentials, denoted by $W_s (x,a_0)$ with $s=0,1,\cdots,k-1$, such that the respective partner potentials $V_s^{(\pm)} (x,a_0)$ defined in Eq. (\ref{Vpm}) obey the following $k$ relations
\ba
V_0^{(+)} (x,a_0) & = & V_1^{(-)} (x,a_0) + R_0(a_0) \, ,
\nonumber
\\
V_1^{(+)} (x,a_0) & = & V_2^{(-)} (x,a_0) + R_1(a_0) \, ,
\nonumber
\\
\cdots & = & \cdots
\nonumber
\\
V_{k-1}^{(+)} (x,a_0) & = & V_0^{(-)} (x,a_1) + R_{k-1}(a_0) \, ,
\la{SIPk}
\ea
where the $k$ arbitrary remainders $R_s(a_0)$ are independent of $x$.
By shape invariance condition in $k$ steps, we mean that the SUSY partner potentials $V_0^{(-)}(x,a_1)$ and $V_0^{(+)}(x,a_0)$ at this time are related to each other by the above $k$ relations.

It is readily shown that from Eq. (\ref{SIPk}) the energy eigenvalues for the initial potential $V_0^{(-)} (x,a_0)$ can be algebraically determined.
These eigenvalues are found as
\be
E_{nk+s}^{(-)} = \sum_{m=0}^{n-1} \sum_{t=0}^{k-1} R_t(a_m) + \sum_{t=0}^{s-1} R_t(a_n) \, ,
\la{Ekn}
\ee
where $E_{0}^{(-)} = 0$, $s=0,1,\cdots,k-1$, and \( n=0,1,2,\cdots \).
Note that we use the convention for the summation: $\sum_{t=0}^{-1}=0$.

The purpose of the present article is to explore the possible algebraic structures of shape invariance condition in $k$ steps, as described by Eqs. (\ref{SIPk}) and (\ref{Ekn}).
The general solution to this problem remains unsolved.
However, as we shall show that if extra relations among the $k$ unrelated superpotentials $W_s (x,a_0)$ and among the $k$ unrelated remainders $R_s (a_0)$ can be separately introduced, the underlying potential algebra will be simplified.
This simplified version of potential algebra is found to be equivalent to the so-called generalized deformed oscillator algebra with a built-in $Z_k$-grading structure.
Therefore, the algebra of shape invariance condition in $k$ steps can be studied in the context of the $Z_k$-graded generalized deformed oscillator \cite{qv,dk}.
For the purpose of illustration, we explicitly work out the detailed algebraic properties of shape invariant potentials in general $k$ steps, in which the parameters of partner potentials are related to each other by a translational change \( a_1=a_0+\delta \).
These results include the cyclic shape invariant potentials of period $k$ as a particular case
\cite{cyclic,qv1,dk1}.
In addition, the results also contain two new types of shape invariant potentials in $k$ steps:
The P\"oschl-Teller I \& II potentials in $k$ steps that generalize the ordinary shape invariant P\"oschl-Teller I \& II potentials from one-step to $k$ steps, respectively.

The article is organized as follows.
In Section \ref{Algebraic shape invariance in $k$ steps}, by imposing the extra relations we establish the simplified version of potential algebra of shape invariance condition in $k$ steps.
In Section \ref{$Z_k$-graded generalized deformed oscillator}, we review and modify the basic definitions and notations of the generalized deformed oscillator with a built-in $Z_k$-grading structure.
In Section \ref{Algebraic structures of translational shape invariant potentials}, we explicitly work out the algebraic properties of translational shape invariant potentials in $k$ steps, based on the $Z_k$-graded deformed oscillator algebra.
Finally, Section \ref{Conclusions} contains a discussion of the present article.


\section{\la{Algebraic shape invariance in $k$ steps}
             Algebraic shape invariance in $\mbox{\boldmath$k$}$ steps}

It is known that shape invariance condition in one-step (\ref{SIP1}) possesses what is generally referred to as a potential algebra \cite{biw,cww}.
That is, the one-step shape invariant potentials have an underlying algebraic structure, thus can be studied by group theoretical methods \cite{gcg,balantekin}.
As a result, the energy eigenvalues of the one-step shape invariant Hamiltonian can be determined by purely algebraic means.

With some restriction, the concept of potential algebra can be applied to the shape invariance condition in more than one-step, too.
Recently, the simplified version of potential algebra for shape invariance condition in two steps has been constructed and discussed by the author \cite{su2}.
It is found that the corresponding simplified potential algebra is similar to that of shape invariance in one-step, and is based only on the three angular-momentum-like generators.
It is also suggested that the same technique can be further extended to the shape invariance condition in more than two steps.
In this section, we shall continue to investigate the general algebraic properties for the shape invariance condition in multi-steps in a concrete and consistent way.

To begin with, let us consider in Eq. (\ref{SIPk}) the substitution of parameters
$a_0 \to \mbox{\boldmath$\alpha$}(N_0)$ and for the general case the substitution
$a_m \to \mbox{\boldmath$\alpha$}(N_0-m)$, where $N_0$ is an arbitrary positive integer and $m=0,1,2,\cdots$.
The precise form of the function $\mbox{\boldmath$\alpha$}(N_0)$ is to be determined by requiring that the change $\mbox{\boldmath$\alpha$}(N_0) \to \mbox{\boldmath$\alpha$}(N_0-1)$ corresponds to the change of parameters $a_0 \to a_1$.
In terms of the $k$ superpotentials, $W_s(x,\mbox{\boldmath$\alpha$}(N_0))$ with
\( s=0,1,\cdots,k-1 \), the corresponding shape invariance condition in $k$ steps (\ref{SIPk}) can be expressed as
\ba
W_0^2(\mbox{\boldmath$\alpha$}(N_0)) + W_0'(\mbox{\boldmath$\alpha$}(N_0)) &=&
W_1^2(\mbox{\boldmath$\alpha$}(N_0)) - W_1'(\mbox{\boldmath$\alpha$}(N_0)) +
R_0(\mbox{\boldmath$\alpha$}(N_0)) \, ,
\nonumber
\\
W_1^2(\mbox{\boldmath$\alpha$}(N_0)) + W_1'(\mbox{\boldmath$\alpha$}(N_0)) &=&
W_2^2(\mbox{\boldmath$\alpha$}(N_0)) - W_2'(\mbox{\boldmath$\alpha$}(N_0)) +
R_1(\mbox{\boldmath$\alpha$}(N_0)) \, ,
\nonumber
\\
\cdots &=& \cdots
\la{SIPk.1}
\\
W_{k-1}^2(\mbox{\boldmath$\alpha$}(N_0)) + W_{k-1}'(\mbox{\boldmath$\alpha$}(N_0)) &=&
W_0^2(\mbox{\boldmath$\alpha$}(N_0-1)) - W_0'(\mbox{\boldmath$\alpha$}(N_0-1)) +
R_{k-1}(\mbox{\boldmath$\alpha$}(N_0)) \, .
\nonumber
\ea
To simplify the notation we have suppressed the $x$-dependence in the superpotentials
$W_s(x,\mbox{\boldmath$\alpha$}(N_0))$ in Eq. (\ref{SIPk.1}).
These $k$ superpotentials are quite arbitrary at this stage.
Without further clues, obviously it is very difficult to determine the underlying potential algebra described by the above $k$ relations.

To solve this problem, we restrict ourselves to a particular subset of shape invariance condition in $k$ steps, instead.
The subset is obtained by imposing extra relations on the $k$ superpotentials and on the $k$ remainders.
This will result in a simplified version of the corresponding potential algebra .
Explicitly, these relations are based on the following particular identification:
\ba
W_s (x,\mbox{\boldmath$\alpha$}(N_0)) \equiv
W (x,\mbox{\boldmath$\alpha$}(N_0-{\textstyle\frac{s}{k}})) \, ,
~~~~~
R_s (\mbox{\boldmath$\alpha$}(N_0)) \equiv
R (\mbox{\boldmath$\alpha$}(N_0-{\textstyle\frac{s}{k}})) \, .
\la{identification}
\ea
With the help of Eq. (\ref{identification}), the $k$ seemly unrelated relations in (\ref{SIPk.1}) can be cast into a single and compact equation for the unified superpotential
$W(x,\mbox{\boldmath$\alpha$}(N_0))$ and the unified remainder $R(\mbox{\boldmath$\alpha$}(N_0))$ as
\ba
&& W^2 (x,\mbox{\boldmath$\alpha$}(N_0-{\textstyle\frac{s}{k}})) +
W' (x,\mbox{\boldmath$\alpha$}(N_0-{\textstyle\frac{s}{k}}))
\nonumber \\
&&~~~~~ \, = \,
W^2 (x,\mbox{\boldmath$\alpha$}(N_0-{\textstyle\frac{s+1}{k}})) -
W' (x,\mbox{\boldmath$\alpha$}(N_0-{\textstyle\frac{s+1}{k}})) +
R (\mbox{\boldmath$\alpha$}(N_0-{\textstyle\frac{s}{k}})) \, .
\la{compactW}
\ea
We note that the $k$ relations in Eq. (\ref{SIPk.1}) can be easily reproduced from Eq. (\ref{compactW}) by letting, one at a time, \( s=0,2,\cdots,k-1 \), respectively.

Eq. (\ref{compactW}) actually represents a constraint equation for the superpotential $W(x,\mbox{\boldmath$\alpha$}(N_0))$, when the parameter $N_0$ is changed by $-\frac{1}{k}$ as we go from the first superpotential to the second one, to the third one, to the fourth one, and so on.
In the viewpoint of quantum mechanics, this change of parameter, {\it i.e.},
\( N_0 \to N_0-\frac{1}{k} \), can be formally accomplished by the action of the raising and lowering operators of angular momentum \cite{gcg}, or equivalently, by the action of the creation and annihilation operators of simple harmonic oscillator \cite{dddd}.
Here, we shall adopt the notion of simple harmonic oscillator in the description and treatment of our problem.
With this in mind, we first define an operator ${\cal N}$, that is analogous to the number operator of the harmonic oscillator, by
\be
{\cal N}  \, \equiv \, \frac{1}{i} \frac{\partial}{\partial \phi} \, ,
\la{J3.2}
\ee
and designate the parameter $N_0$ as the eigenvalue of the number operator ${\cal N}$ acting on the eigenstate $\left| N_0 \right>$, that is, \( {\cal N} \left| N_0 \right> = N_0 \left| N_0 \right> \).
Now, it is natural to think that the unified superpotential function lives in a two-dimensional space spanned by the coordinates $x$ and $\phi$, and the eigenvalue of ${\cal N}$ is assigned to the labeling parameter.
In this way, the set of $k$ relations in Eq. (\ref{SIPk.1}) can be immediately generated, when we project this operator equation
\ba
&&
W^2 (x,\mbox{\boldmath$\alpha$}({\cal N}+{\textstyle\frac{1}{k}})) +
W' (x,\mbox{\boldmath$\alpha$}({\cal N}+{\textstyle\frac{1}{k}}))
\nonumber \\
&&~~~~~ \, = \,
W^2(x,\mbox{\boldmath$\alpha$}({\cal N})) -
W'(x,\mbox{\boldmath$\alpha$}({\cal N})) +
R (\mbox{\boldmath$\alpha$}({\cal N}+{\textstyle\frac{1}{k}})) \, ,
\la{WSIP2}
\ea
into the eigenstates $\big| N_0-\frac{s+1}{k} \big>$, for $s=0,1,\cdots,k-1$, respectively.

Next, we construct the annihilation and creation operators of the associated harmonic oscillator.
The annihilation operator ${\cal A}$ and creation operator ${\cal A}^\dagger = ({\cal A})^\dagger$ are built using the unified superpotential \( W(x,\alpha({\cal N})) \) as
\ba
{\cal A} = e^{-i\phi/k} \bigg[ \frac{\partial}{\partial x} +
W \! \left( x, \mbox{\boldmath$\alpha$}({\cal N}) \right) \bigg] \, ,
~~~~~
{\cal A}^\dagger = \bigg[ - \frac{\partial}{\partial x} +
W \! \left( x, \mbox{\boldmath$\alpha$}({\cal N}) \right) \bigg] e^{i\phi/k} \, .
\la{JmJp.2}
\ea
Explicit computations then show that the simplified potential algebra of shape invariance in $k$ steps, based on the set of generators \( \{ \, I, {\cal A}^\dagger, {\cal A}, {\cal N} \, \} \), is indeed closed and generically nonlinear.
Their commutation relations are described by
\be
[ \, {\cal A} , {\cal N} \, ] = \frac{1}{k}{\cal A} \, , ~~~~~~
[ \, {\cal A}^\dagger , {\cal N} \, ] = -\frac{1}{k}{\cal A}^\dagger \, , ~~~~~~
[ \, {\cal A}^\dagger, {\cal A} \, ] \, = \,
- R \Big( \mbox{\boldmath$\alpha$}\Big({\cal N}+\frac{1}{k}\Big)\Big) \, .
\la{algebrak}
\ee
where we have used Eq. (\ref{WSIP2}).
In this way, the operator ${\cal A}^\dagger({\cal A})$ as desired changes the eigenvalues of ${\cal N}$ by $+(-)\frac{1}{k}$, respectively.
We mention here that the above potential algebra (\ref{algebrak}) based on the set of generators
\( \{ \, I, {\cal A}^\dagger, {\cal A}, {\cal N} \, \} \) is similar to that of shape invariance condition in one-step \cite{biw}-\cite{balantekin}.
Moreover, based on our formulation, this algebra is also equivalent to the so-called generalized deformed oscillator algebra, which has been under extensive studied \cite{dddd}-\cite{bjob}.
In other words, the particular identification (\ref{identification}) not only reduces shape invariance condition effectively from $k$ steps to one-step, but also simplifies the associated potential algebra
(\ref{SIPk.1}) to the well-established deformed oscillator algebra.

The representation of the simplified potential algebra (\ref{algebrak}) can be obtained as follows.
We consider the simultaneous eigenstates, denoted by $\big| \frac{n}{k} \big>$ for $n=0,1,2,\cdots$, of the mutually commuting operators ${\cal A}^\dagger {\cal A}$ and ${\cal N}$, whose respective eigenvalue equations are given by
\be
{\cal A}^\dagger{\cal A} \left| \textstyle\frac{n}{k} \right> =
{\cal F} (\mbox{\boldmath$\alpha$}(\textstyle\frac{n}{k}))
\left| \textstyle\frac{n-1}{k} \right> \, , ~~~~~
{\cal N} \left| \textstyle\frac{n}{k} \right> =
\textstyle\frac{n}{k} \left| \textstyle\frac{n}{k} \right> \, ,
\la{eigenequations2}
\ee
respectively.
Here, we shall call ${\cal F} (\mbox{\boldmath$\alpha$}(\frac{n}{k}))$ the structure function as suggested in the generalized deformed oscillator algebra.
Both equations together imply the action of ladder-type operators ${\cal A}$ and ${\cal A}^\dagger$ on the eigenstate $\big| \textstyle\frac{n}{k} \big>$ as
\be
{\cal A} \left| \textstyle\frac{n}{k} \right> =
\sqrt{{\cal F} (\mbox{\boldmath$\alpha$}(\textstyle\frac{n}{k}))} \,
\left| \textstyle\frac{n-1}{k} \right> \, ,
~~~~~
{\cal A}^\dagger \left| \textstyle\frac{n}{k} \right> =
\sqrt{{\cal F} (\mbox{\boldmath$\alpha$}(\textstyle\frac{n+1}{k}))} \,
\left| \textstyle\frac{n+1}{k} \right> \, ,
\ee
where, without loss of generality, the structure function
${\cal F} (\mbox{\boldmath$\alpha$}(\textstyle\frac{n}{k}))$ is chosen to be real and positive.
If the spectrum of the operators ${\cal A}^\dagger{\cal A}$ exhibits a lowest-weight eigenstate, that is, \( {\cal A} \left| 0 \right> = 0 \), we then choose the condition
${\cal F}(\mbox{\boldmath$\alpha$}(0))=0$ to be satisfied.

The energy eigenvalues of the initial Hamiltonian $H_0^{(-)}(x,a_0)$ =
$H_0^{(-)}(x,\mbox{\boldmath$\alpha$}(N_0))$ can be expressed purely in terms of the structure function ${\cal F} (\mbox{\boldmath$\alpha$}(\frac{n}{k}))$.
We determine this relation by simply projecting the third commutator of Eq. (\ref{algebrak}) on the eigenstate $\big| N_0-\frac{s+1}{k} \big>$, and obtain
\be
{\cal F} (\mbox{\boldmath$\alpha$}(N_0-\textstyle\frac{s+1}{k})) -
{\cal F} (\mbox{\boldmath$\alpha$}(N_0-\textstyle\frac{s}{k})) \, = \,
-R(\mbox{\boldmath$\alpha$}(N_0-\textstyle\frac{s}{k})) \, .
\la{Ralpha2}
\ee
Then applying Eq. (\ref{Ralpha2}) recursively, we establish from Eq. (\ref{Ekn}) the eigenenergies of the initial Hamiltonian $H_0^{(-)}(x,\mbox{\boldmath$\alpha$}(N_0))$ (\ref{Hpm}) as
\be
E_{n}^{(-)} \, = \,
\sum_{s=0}^{n-1} R(\mbox{\boldmath$\alpha$}(N_0-\textstyle\frac{s}{k})) \, = \,
{\cal F} (\mbox{\boldmath$\alpha$}(N_0)) -
{\cal F} (\mbox{\boldmath$\alpha$}(N_0-{\textstyle\frac{n}{k}})) \, ,
\la{EnII}
\ee
where \( n=0,1,2,\cdots \).
Eq. (\ref{EnII}) describes the energy spectrum of the simplified shape invariance condition in $k$ steps, as to be compared with the more complicated energy spectrum given in Eq. (\ref{Ekn}).

In the present research, we are primarily interested in the algebraic structures of shape invariant potentials in $k$ steps, in which the parameters of partner potentials are related to each other by translation: \( a_1=a_0+\delta \).
The analytical properties of such translational shape invariant potentials in $k$ steps are of little known, except for the shape invariant potentials in two steps \cite{singular,su1}.
Nevertheless, for the purpose of obtaining their algebraic properties, we are to choose the $k$ remainders $R_s(a_m)$ in Eq. (\ref{SIPk}) having the following particular forms \( (m=0,1,2,\cdots) \)
\be
R_s(a_m) = \sigma_s + a_m \, \omega_s \, ,
\la{Rs}
\ee
where $\sigma_s$ and $\omega_s$, for $s=0,1,\cdots,k-1$, are arbitrary constants.
In fact, the choice for the $k$-step remainders is not unique.
There is other possibility for the remainders, which results in a quadratic polynomial of the parameter $a_m$ \cite{su1}.
However, we will not pursuit this more difficult problem here.

Now, as suggested by the identification (\ref{identification}), the simplified potential algebra associated to shape invariance in $k$ steps can be established if we identify the $k$ unrelated remainders in Eq. (\ref{Rs}) as
\be
R_s(a_m) = R_s(\mbox{\boldmath$\alpha$}(N_0-m)) \equiv
R(\mbox{\boldmath$\alpha$}(N_0-m-{\textstyle\frac{s}{k}})) \, .
\la{Rs.1}
\ee
By comparing both equations (\ref{Rs}) and (\ref{Rs.1}), we immediately find that the unified remainder $R(\mbox{\boldmath$\alpha$}(N_0-m-{\textstyle\frac{s}{k}}))$, having the needed functional property, can be determined if the $k$ pairs of constants $(\sigma_s,\omega_s)$ are related to one another by
\be
\frac{\sigma_s}{\omega_s} = \frac{\sigma_0}{\omega_0} + \frac{s}{k} \, \delta \, .
\la{Rs.2}
\ee
In this way, the unified remainder takes the form
\be
R(\mbox{\boldmath$\alpha$}(N_0-m-{\textstyle\frac{s}{k}})) =
\bigg[ \,
\frac{\sigma_0}{\omega_0} + a_0 + N_0\,\delta - (N_0-m-\textstyle\frac{s}{k})\delta \,
\bigg] \omega_s \, ,
\la{R-unified}
\ee
where the translational change of the parameters \( a_m=a_0+m\delta \) is implied.

A remark is in order at this stage.
The particular form of the unified remainder in Eq. (\ref{R-unified}) actually imposes an extra $Z_k$-grading structure into the energy spectrum of shape invariant potentials in $k$ steps.
It can be easily checked by noting that the unified remainder in Eq. (\ref{R-unified}) is decomposed into $k$ distinct combinations specified by $\omega_s$, for each different choice of $s$
\( (s =0,1,\cdots,k-1) \).
Because of the choice of the remainders (\ref{Rs}), the simplified potential algebra of shape invariance in $k$ steps is further found to be equivalent to the generalized deformed oscillator algebra that has a built-in $Z_{k}$-grading structure.
The algebraic properties of shape invariant potentials in $k$ steps can therefore be formulated by the method of $Z_k$-graded generalized deformed oscillator.
The algebra of $Z_k$-graded deformed oscillator is reviewed and modified in next section.


\section{\la{$Z_k$-graded generalized deformed oscillator}
             $\mbox{\boldmath$Z_k$}$-graded generalized deformed oscillator}

For the purpose to establish the definitions and notations, we shall begin with a brief review on the basic principles of the generalized deformed oscillator algebra.
Then, we introduce the generalized deformed oscillator algebra with a $Z_k$-grading structure.
In order for our presentation to be consistent with the formulations already described in the preceding section, we will make some modifications on the $Z_k$-graded deformed oscillator algebra.

\subsection{{\sl Generalized deformed oscillator algebra}}

Deformed oscillators have been proposed and studied in many different deformation schemes in the literature \cite{dddd}-\cite{bjob}.
Here, we adopt the method of generalized deformed oscillator suggested in \cite{dddd}.
All deformed oscillators can be unified into a common mathematical framework in the formulation of the so-called generalized deformed oscillator.
A generalized deformed oscillator is defined by a nonlinear algebra generated by the operators $I$, $a$, $a^\dagger$, and $N$ that fulfill the Hermiticity conditions $(a)^\dagger=a^\dagger$, $N^\dagger=N$, and the relations
\be
[ a, N ] = a \, , ~~~~~
[ a^\dagger, N] = -a^\dagger \, , ~~~~~
a^\dagger a = F(N) \, , ~~~~~
a a^\dagger = F(N+1) \, , ~~~~~
\la{algebra}
\ee
where $N$ is the number operator and the Hermitian nonnegative function $F(N)$ is the structure function.
In order to have the Fock representation, $F(N)$ should satisfy the condition: $F(0)=0$ and $F(n)>0$, for $n=1,2,3\cdots$.
Technically, the structure function $F(N)$ is characteristic of the deformation scheme.
Different structure function corresponds to different deformed oscillator.
Nevertheless, all the deformed oscillators can be accommodated within the framework as given in Eq. (\ref{algebra}).

To realize the algebra of the generalized deformed oscillator, it is natural to introduce the Fock space of eigenstates of the number operator $N$, which have the property: \( N \left| n \right> = n \left| n \right> \) and $\left< n | m \right> = \delta_{n,m}$ (for $n,m = 0,1,2,\cdots$).
If the ground state satisfies the relation \( a \left| 0 \right> = 0 \), the complete number eigenstates are obtained by successive application of the operator $a^\dagger$ as
\be
\left| n \right> = \frac{1}{\sqrt{F(n)!}} (a^\dagger)^n \left| 0 \right> \, ,
\ee
where $n=0,1,\cdots,d-1$ and $d$ may be finite or infinite.
The normalization coefficients $F(n)!$ are given by
\be
F(n)! = \prod_{k=1}^n F(k) \, , ~~~~~ F(0)! = 1 \, .
\ee
In this sense, the operators $a$ and $a^\dagger$ are the annihilation and creation operators of this deformed oscillator algebra,
\be
a \left| n \right> = \sqrt{F(n)} \left| n-1 \right> \, , ~~~~~
a^\dagger \left| n \right> = \sqrt{F(n+1)} \left| n+1 \right> \, .
\ee

\subsection{{\sl $Z_k$-graded deformed oscillator algebra}}

We introduce the $Z_k$-graded deformed oscillator algebra that possesses a built-in $Z_k$-grading structure in the subsection.
In the literature, there are various versions of $Z_k$-extended deformed oscillator algebra that have been discussed \cite{qv,dk}.
Here, we make some modifications on this $Z_k$-extended algebra for the purpose to incorporate with the algebraic properties of $k$-step shape invariance condition.

To be more specific, the algebra of the $Z_k$-graded generalized deformed oscillator is defined by the operators $I$, $a$, $a^\dagger$, $N$, and $T$ that fulfill the Hermiticity conditions $(a)^\dagger=a^\dagger$, $N^\dagger=N$, $T^\dagger=T^{-1}$ and the following relations
\be
[ a, N ] = \frac{1}{k} \, a \, , ~~~~~
[ a^\dagger, N] = -\frac{1}{k} \, a^\dagger \, , ~~~~~
a^\dagger \, a = F(N) \, , ~~~~~
a \, a^\dagger = F\big(N+\textstyle\frac{1}{k}\big) \, ,
\la{algebra.k1}
\ee
\be
T^k = I \, , ~~~~~
[ N, T ] = 0 = [ N, T^\dagger ] \, , ~~~~~
a^\dagger \, T = e^{-{\rm i}2\pi/k} \, T \, a^\dagger \, , ~~~~~
T^\dagger \, a = a \, T^\dagger \, e^{{\rm i}2\pi/k} \, ,
\la{algebra.k2}
\ee
where $k=1,2,3,\cdots$.
For the special value $k=1$, it is readily known that the algebra defined in Eqs. (\ref{algebra.k1}) and (\ref{algebra.k2}) reduces to the generalized deformed oscillator algebra (\ref{algebra}).
In addition, we note that in defining the $Z_k$-graded deformed oscillator algebra above, the creation operator $a^\dagger$ is designated to increase the eigenvalue of the number operator $N$ by $\frac{1}{k}$ unit, not by unity as in the conventional $Z_k$-extended deformed oscillator.
Similarly, the annihilation operator $a$ decreases that by $\frac{1}{k}$ unit.

The grading operator $T$ in Eq. (\ref{algebra.k2}) is the generator of the cyclic group of order $k$,
\( Z_k = \{ 1,T,T^2,\cdots,T^{k-1} | T^k=1 \} \).
There can be many different realizations for such a operator, but here, we choose the one, in which the grading operator is easily realized in terms of the number operator $N$ as
\be
T = e^{2\pi {\rm i} N} \, .
\ee
Besides, the grading operator $T$ has $k$ distinct eigenvalues $q^s$, in which $q$ is the $s$-th root of unity and $s=0,1,2,\cdots,k-1$.
Therefore, the $Z_k$-grading structure of the Fock space ${\cal H}$ can be distinguished by the  eigenvalues of the grading operator $T$, for which the corresponding Fock subspace of eigenvalue $q^s$ is denoted by
\be
{\cal H}_{s} :
\left\{ \, \left| n+\textstyle\frac{s}{k} \right> \Big| \, n=0,1,2,\cdots \, \right\} \, .
\ee
The entire Fock space is consequently the direct sum of each individual Fock subspace as given by
\( {\cal H} = \sum_{s=0}^{k-1} \oplus {\cal H}_{s} \).

We have shown that the Fock space of the $Z_k$-graded deformed oscillator consists of $k$ distinct Fock subspaces, which can be specified by the grading operator $T$.
In fact, the grading structure of the Fock space can also be naturally formulated by the projection operators that by construction project on these distinct subspaces.
Explicitly, the projection operators are expressed in terms of the grading operator $T$ as
(for \( s=0,1,\cdots,k-1 \))
\be
\Pi_s = \frac{1}{k} \sum_{t=0}^{k-1}  e^{-2\pi {\rm i} ts/k} T^{\, t} =
\frac{1}{k} \sum_{t=0}^{k-1} e^{2\pi {\rm i} t(N-s/k)} \, , ~~~~~
\sum_{s=0}^{k-1} \Pi_s = I \, .
\la{projection}
\ee

At this stage, we are able to restate the algebra of the $Z_k$-graded generalized deformed oscillator, using the projection operator $\Pi_s$, not by the grading one $T$:
The algebra of $Z_k$-graded deformed oscillator is generated by the set of operators $I$, $a$, $a^\dagger$, $N$, and $\Pi_s$, that fulfill the Hermiticity conditions $(a)^\dagger=a^\dagger$, $N^\dagger=N$, $\Pi_s^\dagger =\Pi_s$ and the following relations
\be
[ a, N ] = \frac{1}{k} \, a \, , ~~~~~
[ a^\dagger, N ] = -\frac{1}{k} \, a^\dagger \, , ~~~~~
a^\dagger \, a = F(N) \, , ~~~~~
a \, a^\dagger = F\big(N+\textstyle\frac{1}{k}\big) \, ,
\la{algebra.k3}
\ee
\be
[ N, \Pi_s ] = 0 \, , ~~~~~
\Pi_s \Pi_t = \delta_{s,t} \, , ~~~~~
a^\dagger \, \Pi_s = \Pi_{s+1} \, a^\dagger \, , ~~~~~
a \, \Pi_s = \Pi_{s-1} \, a \, ,
\la{algebra.k4}
\ee
where $s,t=0,1,\cdots,k-1$.
The convention for the projection operators has been used: \( \Pi_{t} = \Pi_s \), if \( t-s=0 \) mod $k$.

To establish the complete Fock space representation for the $Z_k$-graded deformed oscillator, let us again use the creation and annihilation operators $a^\dagger$ and $a$, and take the eigenstates of the number operator $N$, which fulfill the properties:
\( N \big| \frac{n}{k} \big> = \frac{n}{k} \big| \frac{n}{k} \big> \) and
$\big< \frac{n}{k} | \frac{m}{k} \big> = \delta_{n,m}$, for $n,m = 0,1,2,\cdots$.
The ground state is defined by the simultaneous eigenstate of the operators $N$ and $a^\dagger a$ with both vanishing eigenvalues, and in addition satisfies the relation \( a \left| 0 \right> = 0 \).
The complete Fock-space eigenstates can be then constructed by successive application of the creation operator $a^\dagger$ as
\be
\left| \textstyle\frac{n}{k} \right> =
\frac{1}{\sqrt{F\big(\textstyle\frac{n}{k}\big)!}} (a^\dagger)^n \Big| 0 \Big> \, ,
\ee
where the normalization coefficients are
\be
F \big({\textstyle\frac{n}{k}}\big)! = \prod_{m=1}^n F\big({\textstyle\frac{m}{k}}\big) \, ,
~~~~~ F_k(0)! = 1 \, .
\ee
In this way, the eigenvalue equations for the operators $N$ and $\Pi_t$ are
\be
N \left| n+\textstyle\frac{s}{k} \right> =
(n+\textstyle\frac{s}{k}) \left| n+\textstyle\frac{s}{k} \right> \, ,
\ee
\be
\Pi_{t} \left| n+\textstyle\frac{s}{k} \right> =
\delta_{t,s} \left| n+\textstyle\frac{s}{k} \right> \, ,
\ee
where $s,t=0,1,\cdots,k-1$.
Moreover, the operators $a$ and $a^\dagger$ act on these eigenstates rendering
\be
a \left| n+\textstyle\frac{s}{k} \right> =
\sqrt{F\big(n+\textstyle\frac{s}{k}\big)} \left| n+\textstyle\frac{s-1}{k} \right> \, ,
\ee
\be
a^\dagger \left| n+\textstyle\frac{s}{k} \right> =
\sqrt{F\big(n+\textstyle\frac{s+1}{k}\big)} \left| n+\textstyle\frac{s+1}{k} \right> \, .
\ee

Finally, because of the encoded $Z_k$-grading structure, the corresponding structure function $F(N)$ will have the following form, expressed in terms of the projection operators $\Pi_s$, as
\be
F(N) = f(N) + \sum_{s=0}^{k-1} g_s(N) \, \Pi_s \, .
\la{FN}
\ee
Here, $f(N)$ and $g_s(N)$ are analytic functions of $N$.


\section{\la{Algebraic structures of translational shape invariant potentials}
             Algebraic structures of translational shape invariant potentials}

To illustrate the simplified potential algebra developed in Section \ref{Algebraic shape invariance in $k$ steps}, we shall investigate algebraic structures for the shape invariant potentials in $k$ steps, in which the parameters of partner potentials are related to each other by a translation change: $a_m$ = $a_{m-1}+\delta$ = $a_0 + m\delta$.
The quantity $\delta$ is a constant and \( m=0,1,2,\cdots \).
The analysis will be based on the formalism of the $Z_k$-graded deformed oscillator reviewed and modified in Section \ref{$Z_k$-graded generalized deformed oscillator}.
The relevant algebraic quantities for the shape invariant potentials in the first few number of $k$ steps will be explicitly constructed.
After getting enough information for these potentials, the general algebraic results for the shape invariant potentials in arbitrary $k$ steps can be deduced.

Before presenting the detailed parts, we mention here some important relations that are used in the upcoming calculations.
We note that the unified remainder given in Eq. (\ref{R-unified}) can be expressed in terms of the number operator ${\cal N}$ and projection operators $\Pi_s$.
It takes the compact form as
\be
R(\mbox{\boldmath$\alpha$}({\cal N})) =
\Big( {\cal C}-(N_0-{\cal N})\delta \Big) \sum_{s=0}^{k-1} \omega_s \Pi_s \, ,
\la{R-compact}
\ee
where \( {\cal C} = \frac{\sigma_0}{\omega_0} + a_0 \) is a short hand notation .
In the same vein, using the operators ${\cal N}$, $\Pi_s$, and Eq. (\ref{FN}), we are able to write the structure function ${\cal F} (\mbox{\boldmath$\alpha$}({\cal N}))$ of shape invariance condition in $k$ steps defined in Eq. (\ref{eigenequations2}) as
\be
{\cal F}(\mbox{\boldmath$\alpha$}({\cal N})) =
f({\cal N}) + \sum_{s=0}^{k-1} g_s({\cal N}) \, \Pi_s \, ,
\la{F-compact}
\ee
where $f({\cal N})$ and $g_s({\cal N})$ are functions of ${\cal N}$.
All of them are determined by requiring Eq. (\ref{F-compact}) to satisfy the remainder-structure function relation (\ref{Ralpha2}), that is,
\be
{\cal F}(\mbox{\boldmath$\alpha$}({\cal N}-\textstyle\frac{1}{k})) -
{\cal F}(\mbox{\boldmath$\alpha$}({\cal N})) \, = \,
-R(\mbox{\boldmath$\alpha$}({\cal N})) \, .
\la{FFR}
\ee

Furthermore, we introduce two quantities that will be useful in the later presentations.
The first one is the analogous Kronecker delta for the cyclic group of order $k$.
We use the symbol $\Delta_{s,t}$ that is defined by
\be
\Delta_{s,t} =
\left\{
\begin{array}{lc}
1 \, , & ~~{\rm for}~~s = t~~{\rm mod}~~k \, , \\
0 \, , & ~~{\rm for}~~s \ne t~~{\rm mod}~~k \, .
\end{array}
\right.
\la{Delta}
\ee
The second one is the quantity $\Omega_k$ given by the summation
\be
\Omega_k \equiv \sum_{s=0}^{k-1} \omega_s \, .
\ee

\subsection{{\sl Shape invariance in two steps}}

Let us begin with the simplest example, which is the shape invariance condition in two steps.
For $k=2$, the grading operator $\exp(2\pi{\rm i}{\cal N})$ simply reduces to the usual Klein operator.
The corresponding projection operators
$\Pi_0 = \frac{1}{2} (I+(-1)^{2{\cal N}})$ and $\Pi_1 = \frac{1}{2} (I-(-1)^{2{\cal N}})$
project upon the even subspace
\( {\cal H}_0 = \big\{ \big| N_0-n \big> | n=0,1,2,\cdots \, \big\} \) and odd subspace
\( {\cal H}_1 = \big\{ \big| N_0-n-\frac{1}{2} \big> | n=0,1,2,\cdots \, \big\} \) of the
$Z_2$-graded Fock space ${\cal H}$, respectively.
In this sense, the $Z_2$-graded deformed oscillator algebra is similar to the Calogero-Vasiliev oscillator algebra \cite{vasiliev}.

When applying the operator equation (\ref{R-compact}) on the two eigenstates $\big| N_0 \big>$ and $\big| N_0-\frac{1}{2} \big>$, we obtain the first two remainders for the initial Hamiltonian
$H_0^{(-)}(x,\mbox{\boldmath$\alpha$}(N_0))$ as $R(\mbox{\boldmath$\alpha$}(N_0)) = {\cal C}\,\omega_0$ and $R(\mbox{\boldmath$\alpha$}(N_0-\frac{1}{2})) = ({\cal C}+\frac{1}{2}\delta)\omega_1$, respectively.
The values of $R(\mbox{\boldmath$\alpha$}(N_0-\frac{1}{2}n))$ for other number eigenstates can be obtained in the similar manner.
Using these results, the corresponding structure function ${\cal F}(\mbox{\boldmath$\alpha$}({\cal N}))$ for the shape invariant potentials in two steps can be determined by Eq. (\ref{FFR}).
After some calculations, we arrive at
\ba
{\cal F}(\mbox{\boldmath$\alpha$}(N_0-{\textstyle\frac{1}{2}n}))
&=& {\cal C}_0 -
{\textstyle\frac{1}{2}}\,\Omega_2 \big( {\cal C}+{\textstyle\frac{1}{4}}(n-1)\delta \big) \, n +
\nonumber
\\
&& \sum_{s=0}^1
\Big(
\big( {\cal C}+{\textstyle\frac{1}{4}}(2n-1)\delta \big) c_s - \delta \, d_s
\Big) \, \Delta_{s,n} \, ,
\ea
where $n=0,1,2,\cdots$ and ${\cal C}_0$ is an arbitrary constant to render
${\cal F}(\mbox{\boldmath$\alpha$}(N_0-{\textstyle\frac{n}{2}}))$ positive definite.
Due to the presence of $\Delta_{s,n}$ (\ref{Delta}), only one term  in the summation is singled out which fulfills the condition: \( n-s=0 \) mod $2$.
Moreover, the constants $c_s$ and $d_s$ $(s=0,1)$ are given by
\be
c_0 = -c_1 = \textstyle\frac{1}{4}(\omega_0-\omega_1) \, ,~~~~~
d_0 = d_1 = \textstyle\frac{1}{8}(\omega_0+\omega_1) \, .
\ee
The energy spectrum (\ref{EnII}) of the initial Hamiltonian $H_0^{(-)}(x,\mbox{\boldmath$\alpha$}(N_0))$ is therefore
\be
E_{n}^{(-)} \, = \,
{\cal F} (\mbox{\boldmath$\alpha$}(N_0)) -
{\cal F} (\mbox{\boldmath$\alpha$}(N_0-{\textstyle\frac{1}{2}}n)) \, .
\ee
We note that a similar result for shape invariant potentials in two steps has been constructed in \cite{su2} using different approach.

\subsection{{\sl Shape invariance in three steps}}

The next allowed value is $k=3$, that is, the potential algebra of shape invariance condition in three steps.
The grading operator is again denoted by $\exp(2\pi{\rm i}{\cal N})$.
The $Z_3$-graded Fock space is constructed by the direct sum:
${\cal H}=\sum_{s=0}^2 \oplus {\cal H}_s$.
The projection operators are given by
\ba
\Pi_0 &=& \textstyle\frac{1}{3} \big( 1+2\cos{2\pi{\cal N}} \big) \, ,
\nonumber
\\
\Pi_1 &=& \textstyle\frac{1}{3} \big( 1-\cos{2\pi{\cal N}}-\sqrt{3}\,\sin{2\pi{\cal N}} \big) \, ,
\nonumber
\\
\Pi_2 &=& \textstyle\frac{1}{3} \big( 1-\cos{2\pi{\cal N}}+\sqrt{3}\,\sin{2\pi{\cal N}} \big) \, ,
\ea
that project on the Fock subspace
\( {\cal H}_s = \big\{ \big| N_0-n-\frac{s}{3} \big> | n=0,1,2,\cdots \, \big\} \), for $s$ = 0, 1, and 2, respectively.

It is readily checked that when acting Eq. (\ref{R-compact}) on the first three eigenstates
$\big| N_0 \big>$, $\big| N_0-\frac{1}{3} \big>$, and $\big| N_0-\frac{2}{3} \big>$, we obtain
$R(\mbox{\boldmath$\alpha$}(N_0)) = {\cal C}\,\omega_0$,
$R(\mbox{\boldmath$\alpha$}(N_0-\frac{1}{3})) = ({\cal C}+\frac{1}{3}\delta)\omega_1$, and
$R(\mbox{\boldmath$\alpha$}(N_0-\frac{2}{3})) = ({\cal C}+\frac{2}{3}\delta)\omega_2$, respectively.
Other results can be similarly obtained.
By Eq. (\ref{FFR}), we establish the corresponding structure function after some calculations $(n=0,1,2,\cdots)$
\ba
{\cal F}(\mbox{\boldmath$\alpha$}(N_0-{\textstyle\frac{1}{3}n}))
&=& {\cal C}_0 -
{\textstyle\frac{1}{3}}\,\Omega_3 \big( {\cal C}+{\textstyle\frac{1}{6}}(n-1)\delta \big) \, n +
\nonumber
\\
&& \sum_{s=0}^2
\Big(
\big( {\cal C}+{\textstyle\frac{1}{6}}(2n-1)\delta \big) c_s - \delta \, d_s
\Big) \, \Delta_{s,n} \, ,
\la{F3}
\ea
where, as in the previous case, only one term is singled out in the summation that satisfies the condition: \( n-s=0 \) mod $3$.
The constant ${\cal C}_0$ is chosen to make the associated structure function positive definite.
The constants $c_s$ and $d_s$ for $s=0,1,2$ in Eq. (\ref{F3}) are found to be
\ba
c_0 = \textstyle\frac{1}{3}(\omega_0-\omega_2) \, , &&
d_0 = \textstyle\frac{1}{18}(2\omega_0+\omega_1+2\omega_2) \, ,
\nonumber
\\
c_1 = \textstyle\frac{1}{3}(\omega_1-\omega_0) \, , &&
d_1 = \textstyle\frac{1}{18}(2\omega_1+\omega_2+2\omega_0) \, ,
\nonumber
\\
c_2 = \textstyle\frac{1}{3}(\omega_2-\omega_1) \, , &&
d_2 = \textstyle\frac{1}{18}(2\omega_2+\omega_0+2\omega_1) \, .
\ea
The energy spectrum for the shape invariant potentials in three steps can be algebraically determined from Eqs. (\ref{F3}) and (\ref{EnII}), in which $k$ is replaced by $3$.

\subsection{{\sl Shape invariance in four steps}}

The grading operator for shape invariant potentials is $\exp(2\pi{\rm i}{\cal N})$ once more.
The $Z_4$-graded Fock space of the shape invariance condition in four steps is denoted by
${\cal H}=\sum_{s=0}^3 \oplus {\cal H}_s$.
The projection operators are
\ba
\Pi_0 &=& \textstyle\frac{1}{4} \big( 1+2\cos{2\pi{\cal N}}+\cos{4\pi{\cal N}} \big) \, ,
\nonumber
\\
\Pi_1 &=& \textstyle\frac{1}{4} \big( 1-2\sin{2\pi{\cal N}}-\cos{4\pi{\cal N}} \big) \, ,
\nonumber
\\
\Pi_2 &=& \textstyle\frac{1}{4} \big( 1-2\cos{2\pi{\cal N}}+\cos{4\pi{\cal N}} \big) \, ,
\nonumber
\\
\Pi_3 &=& \textstyle\frac{1}{4} \big( 1+2\sin{2\pi{\cal N}}-\cos{4\pi{\cal N}} \big) \, ,
\ea
projecting on the Fock subspace
\( {\cal H}_s = \big\{ \big| N_0-n-\frac{s}{4} \big> | n=0,1,2,\cdots \, \big\} \), for $s=0,1,2$, and 3, respectively.
When acting Eq. (\ref{R-compact}) on the first four eigenstates of the system
$\big| N_0 \big>$, $\big| N_0-\frac{1}{4} \big>$, $\big| N_0-\frac{2}{4} \big>$, and
$\big| N_0-\frac{3}{4} \big>$, we find that
$R(\mbox{\boldmath$\alpha$}(N_0)) = {\cal C}\,\omega_0$,
$R(\mbox{\boldmath$\alpha$}(N_0-\frac{1}{4})) = ({\cal C}+\frac{1}{4}\delta)\omega_1$,
$R(\mbox{\boldmath$\alpha$}(N_0-\frac{2}{4})) = ({\cal C}+\frac{2}{4}\delta)\omega_2$, and
$R(\mbox{\boldmath$\alpha$}(N_0-\frac{3}{4})) = ({\cal C}+\frac{3}{4}\delta)\omega_3$, respectively.
Other values of $R(\mbox{\boldmath$\alpha$}(N_0-\frac{1}{4}n))$ can be similarly constructed.
From Eq. (\ref{FFR}), the structure function can be determined
\ba
{\cal F}(\mbox{\boldmath$\alpha$}(N_0-{\textstyle\frac{1}{4}n}))
&=& {\cal C}_0 -
{\textstyle\frac{1}{4}}\,\Omega_4 \big( {\cal C}+{\textstyle\frac{1}{8}}(n-1)\delta \big) \, n +
\nonumber
\\
&& \sum_{s=0}^3
\Big(
\big( {\cal C}+{\textstyle\frac{1}{8}}(2n-1)\delta \big) c_s - \delta \, d_s
\Big) \, \Delta_{s,n} \, ,
\la{F4}
\ea
where $n=0,1,2,\cdots$ and ${\cal C}_0$ is chosen to set the structure function positive definite.
In the above summation, only the term, satisfying the condition: \( s=n \) mod $4$, contributes to the result.
The constants $c_s$ and $d_s$ for $s=0,1,2,3$ are given by
\ba
c_0 = \textstyle\frac{1}{8}(3\omega_0+\omega_1-\omega_2-3\omega_3) \, , &&
d_0 = \textstyle\frac{1}{32}(3\omega_0+\omega_1+\omega_2+3\omega_3) \, ,
\nonumber
\\
c_1 = \textstyle\frac{1}{8}(3\omega_1+\omega_2-\omega_3-3\omega_0) \, , &&
d_1 = \textstyle\frac{1}{32}(3\omega_1+\omega_2+\omega_3+3\omega_0) \, ,
\nonumber
\\
c_2 = \textstyle\frac{1}{8}(3\omega_2+\omega_3-\omega_0-3\omega_1) \, , &&
d_2 = \textstyle\frac{1}{32}(3\omega_2+\omega_3+\omega_0+3\omega_1) \, ,
\nonumber
\\
c_3 = \textstyle\frac{1}{8}(3\omega_3+\omega_0-\omega_1-3\omega_2) \, , &&
d_3 = \textstyle\frac{1}{32}(3\omega_3+\omega_0+\omega_1+3\omega_2) \, .
\ea
The energy spectrum for the shape invariant potentials in four steps can be algebraically determined from Eqs. (\ref{F4}) and (\ref{EnII}), where $4$ replaces $k$ in the latter equation.

\subsection{{\sl Shape invariance in five steps}}

Before arriving at the discussion on the shape invariance condition in arbitrary $k$ steps, let us present one more example, that is, $k=5$.
After this example, we should gather enough information and will be able to deduce the common algebraic structures that share with all the translational shape invariant potentials.

The $Z_5$-graded Fock space for the shape invariance in five steps is ${\cal H}=\sum_{s=0}^4 \oplus {\cal H}_s$.
The Fock subspaces are
\( {\cal H}_s = \big\{ \big| N_0-n-\frac{s}{5} \big> | n=0,1,2,\cdots \, \big\} \), for $s=0,1,2,3,4$.
The projection operators are given by
\ba
\Pi_0 &=& \textstyle\frac{1}{5} \big( 1+2\cos{2\pi {\cal N}}+2\cos{4\pi {\cal N}} \big) \, ,
\nonumber
\\
\Pi_1 &=& \textstyle\frac{1}{5}
\big( 1-c_2\sin{2\pi{\cal N}}-c_4\cos{4\pi{\cal N}}-s_2\sin{2\pi{\cal N}}+s_4\sin{4\pi{\cal N}} \big) \, ,
\nonumber
\\
\Pi_2 &=& \textstyle\frac{1}{5}
\big( 1-c_4\sin{2\pi{\cal N}}-c_2\cos{4\pi{\cal N}}+s_4\sin{2\pi{\cal N}}+s_2\sin{4\pi{\cal N}} \big) \, ,
\nonumber
\\
\Pi_3 &=& \textstyle\frac{1}{5}
\big( 1-c_4\sin{2\pi{\cal N}}-c_2\cos{4\pi{\cal N}}-s_4\sin{2\pi{\cal N}}-s_2\sin{4\pi{\cal N}} \big) \, ,
\nonumber
\\
\Pi_4 &=& \textstyle\frac{1}{5}
\big( 1-c_2\sin{2\pi{\cal N}}-c_4\cos{4\pi{\cal N}}+s_2\sin{2\pi{\cal N}}-s_4\sin{4\pi{\cal N}} \big) \, ,
\ea
where $c_2=\frac{1}{2}(1-\sqrt{5})$, $c_4=\frac{1}{2}(1+\sqrt{5})$,
$s_2=\sqrt{(5+\sqrt{5})/2}$, and $s_4=-\sqrt{(5-\sqrt{5})/2}\,$.

When the remainder $R(\mbox{\boldmath$\alpha$}({\cal N}))$ in Eq. (\ref{R-compact}) acts on the first five eigenstates $\big| N_0 \big>$, $\big| N_0-\frac{1}{5} \big>$, $\big| N_0-\frac{2}{5} \big>$,
$\big| N_0-\frac{3}{5} \big>$, and $\big| N_0-\frac{4}{5} \big>$,
the corresponding eigenvalues are ${\cal C}\,\omega_0$, $({\cal C}+\frac{1}{5}\delta)\omega_1$,
$({\cal C}+\frac{2}{5}\delta)\omega_2$, $({\cal C}+\frac{3}{5}\delta)\omega_3$, and
$({\cal C}+\frac{4}{5}\delta)\omega_4$, respectively.
Other values can be obtained in the same way.
The structure function for the shape invariant potentials in five steps is determined by Eq. (\ref{FFR}) as $(n=0,1,2,\cdots)$
\ba
{\cal F}(\mbox{\boldmath$\alpha$}(N_0-{\textstyle\frac{1}{5}n}))
&=& {\cal C}_0 -
{\textstyle\frac{1}{5}}\,\Omega_5 \big( {\cal C}+{\textstyle\frac{1}{10}}(n-1)\delta \big) \, n +
\nonumber
\\
&& \sum_{s=0}^4
\Big(
\big( {\cal C}+{\textstyle\frac{1}{10}}(2n-1)\delta \big) c_s - \delta \, d_s
\Big) \, \Delta_{s,n} \, .
\la{F5}
\ea
Similarly, the term, satisfying the condition \( s=n \) mod $5$, survives in the above summation.
The constant ${\cal C}_0$ is chosen to make the associated structure function positive.
The constants $c_s$ and $d_s$ for $s=0,1,2,3,4$ are given in the following structural patterns
\ba
c_0 = \textstyle\frac{1}{5}(2\omega_0+\omega_1-\omega_3-2\omega_4) \, , &&
d_0 = \textstyle\frac{1}{50}(4\omega_0+\omega_1+\omega_3+4\omega_4) \, ,
\nonumber
\\
c_1 = \textstyle\frac{1}{5}(2\omega_1+\omega_2-\omega_4-2\omega_0) \, , &&
d_1 = \textstyle\frac{1}{50}(4\omega_1+\omega_2+\omega_4+4\omega_0) \, ,
\nonumber
\\
c_2 = \textstyle\frac{1}{5}(2\omega_2+\omega_3-\omega_0-2\omega_1) \, , &&
d_2 = \textstyle\frac{1}{50}(4\omega_2+\omega_3+\omega_0+4\omega_1) \, ,
\nonumber
\\
c_3 = \textstyle\frac{1}{5}(2\omega_3+\omega_4-\omega_1-2\omega_2) \, , &&
d_3 = \textstyle\frac{1}{50}(4\omega_3+\omega_4+\omega_1+4\omega_2) \, ,
\nonumber
\\
c_4 = \textstyle\frac{1}{5}(2\omega_4+\omega_0-\omega_2-2\omega_3) \, , &&
d_4 = \textstyle\frac{1}{50}(4\omega_4+\omega_0+\omega_2+4\omega_3) \, .
\ea
The energy spectrum for the shape invariant potentials in five steps can be algebraically determined from Eqs. (\ref{F5}) and (\ref{EnII}).

\subsection{{\sl Shape invariance in arbitrary $k$ steps}}

We are now at a position to deduce the generally algebraic structures for the shape invariant potentials in arbitrary $k$ steps, based on the systematic properties of the previous four examples shown above.

The $Z_k$-graded Fock space of the shape invariance in $k$ steps is ${\cal H}=\sum_{s=0}^{k-1} \oplus {\cal H}_s$.
The Fock subspaces are denoted by
\( {\cal H}_s=\big\{\big| N_0-n-\frac{s}{k} \big> | n=0,1,2,\cdots\,\big\} \), for $s=0,1,\cdots,k-1$.
The remainder $R(\mbox{\boldmath$\alpha$}({\cal N}))$ (\ref{R-compact}) acts on the eigenstates $\big| N_0-\frac{n}{k} \big>$ rendering this result:
$({\cal C}+\frac{n}{k}\delta)\sum_{s=0}^{k-1}\omega_s \Delta_{s,n}$.
The structure function for the shape invariant potentials in $k$ steps is again constructed by Eq. (\ref{FFR}) as $(n=0,1,2,\cdots)$
\ba
{\cal F}\Big(\mbox{\boldmath$\alpha$}\Big(N_0-\frac{n}{k}\Big)\Big)
&=& {\cal C}_0 -
\frac{\Omega_k}{k}\,\Big( {\cal C}+\frac{1}{2k}(n-1)\delta \Big) \, n +
\nonumber
\\
&& \sum_{s=0}^{k-1}
\bigg[
\Big( {\cal C}+\frac{1}{2k}(2n-1)\delta \Big) c_s - \delta \, d_s
\bigg] \, \Delta_{s,n} \, .
\la{Fk}
\ea
In the summation, the $\Delta_{s,n}$ term chooses what satisfies the condition \( s=n \) mod $k$.
${\cal C}_0$ is a constant to render the associated structure function positive.
Moreover, the constants $c_s$ and $d_s$ for $s=0,1,\cdots,k-1$ are determined based on the structural similarities in the corresponding counterparts of the previous four examples.
After some algebras, the general results of $c_s$ and $d_s$ are found to be
\ba
c_s &=& \frac{1}{2k} \sum_{t=0}^{k-1} \, (k-1-2t) \, \omega_{s+t} \, ,
\nonumber
\\
d_s &=& \frac{1}{2k^2} \sum_{t=0}^{k-1} \Big( t^2-(k-1)(t-1) \Big) \, \omega_{s+t} \, ,
\la{csds}
\ea
where we use the convention: \( \omega_{s+t} \equiv \omega_{s+t~{\rm mod}~k} \)
({\it e.g.} $\omega_{k+1} = \omega_{1}$) in these two general formulas.
An interesting observation is that if we define $D(t)=t^2-(k-1)(t-1)$, then
$\frac{{\rm d}}{{\rm d}t}D(t)=-(k-1-2t)$.
It is easily checked that the general form in Eq. (\ref{csds}) exactly reproduces what we have calculated and presented for the shape invariant potentials in the lower steps.
This general form is even true for the higher-step shape invariant potentials in which the detailed results are not shown here.


\section{\la{Conclusions}
             Conclusions}

In the present work, we explore the algebraic properties of shape invariance condition in arbitrary $k$ steps, within the framework of SUSYQM.
By imposing extra relations of identification (\ref{identification}) among the $k$ superpotentials and the $k$ remainders, respectively, we obtain the associated simplified potential algebra of shape invariance in $k$ steps (\ref{algebrak}).
This simplified version of potential algebra is found similar to that of shape invariance in one-step.
Moreover, it is also found to be equivalent to the generalized deformed oscillator algebra, having a built-in $Z_{k}$-grading structure.
As a result, the algebraic realization of shape invariance condition in $k$ steps is formulated by the method of $Z_k$-graded generalized deformed oscillator.
Based on this method, the representation of the simplified potential algebra of shape invariance in $k$ steps can be constructed in terms of the generators of the deformed harmonic oscillator \( \{ I, {\cal A}, {\cal A}^\dagger, {\cal N} \}\) as well as the grading generator $T$ of the cyclic group of order $k$.

For the purpose of illustration, we in addition work out four typical examples of shape invariant potentials in $k=2,3,4,$ and 5 steps, respectively, in which the parameters $a_0$ and $a_1$ of partner potentials are related to each other by translation \( a_1 = a_0 + \delta \).
We here assume that the remainder function depends only linearly on the translational parameter by \( R_s(a_m) = \sigma_s + a_m \, \omega_s \) ($s=0,2,\cdots,k-1$) as given in Eq. (\ref{Rs}).
In each example, the projection operators $\Pi_s$ are explicitly constructed, which project upon the respective Fock subspace ${\cal H}_s$ of the $Z_k$-graded Fock space ${\cal H}$.
The associated structure function ${\cal F}(\mbox{\boldmath$\alpha$}({\cal N}))$ for each example is also established by directly solving Eq. (\ref{FFR}).
Next, by inspecting the algebraic structures for these four examples, we learn that the constants $c_s$ and $d_s$ appearing in the structure function seem to follow some especially systematic rules.
Hence, we continue to explore the shape invariance condition in arbitrary $k$ steps and by deduction establish the general algebraic properties for the translational shape invariant potentials.
The energy eigenvalues of the initial Hamiltonian of shape invariance in $k$ steps consequently can be determined by purely algebraic means.

We conclude the paper by pointing out that the results obtained in this work contains the so-called cyclic shape invariant potentials of period $k$ as a special case.
This can be shown by setting $\delta=0$ in Eq. (\ref{Fk}), then the resultant structure function is nothing but the structure function of the harmonic oscillator potentials in $k$ steps, in which the remainders $R(\mbox{\boldmath$\alpha$}({\cal N}))$ (up to a constant) are arranged as \( \omega_0, \omega_1, \omega_2, \cdots, \omega_{k-1}, \omega_0, \cdots \).
Such potential has been reported in the study of cyclic shape invariant potentials \cite{cyclic,qv1,dk1}.
Moreover, if we set $\delta>0$ in Eq. (\ref{Fk}), a new type of shape invariant potentials in $k$ steps occurs, which can be constructed presumably by generalizing the ordinary shape invariant P\"oschl-Teller I potential from one-step to $k$-steps.
In principle, these P\"oschl-Teller I potentials in $k$ steps possess infinite many number of eigenstates.
Similarly, if we set $\delta>0$ in Eq. (\ref{Fk}), the other type of shape invariant potentials in $k$ steps results.
They are the P\"oschl-Teller II potentials in $k$ steps that generalize the ordinary P\"oschl-Teller II potential from shape invariance in one-step to $k$ steps, and contain a finite number of bound states.
The analytical properties of the P\"oschl-Teller I \& II potentials in two steps have been discussed in Refs. \cite{singular,su1}.

\end{document}